# Valorizing Sewage Sludge: Using Nature-Inspired Architecture to Overcome Intrinsic Weaknesses of Waste-Based Materials

Sabrina C. Shen[1,2], Branden Spitzer[1,2], Damian Stefaniuk[3], Shengfei Zhou[1,4], Admir Masic[3], Markus J. Buehler[1,3,5*]

[1] Laboratory for Atomistic and Molecular Mechanics (LAMM), Massachusetts Institute of Technology, 77 Massachusetts Ave., Cambridge, MA 02139, USA

[2] Department of Materials Science and Engineering, Massachusetts Institute of Technology, 77 Massachusetts Ave., Cambridge, MA 02139, USA

[3] Department of Civil and Environmental Engineering, Massachusetts Institute of Technology, 77 Massachusetts Ave., Cambridge, MA 02139, USA

[4] Wisconsin Energy Institute, Great Lakes Bioenergy Research Center, University of Wisconsin-Madison, 1552 University Ave., Madison, WI 53726, USA

[5] Center for Computational Science and Engineering, Schwarzman College of Computing, Massachusetts Institute of Technology, 77 Massachusetts Ave., Cambridge, MA 02139, USA

*Corresponding author, mbuehler@mit.edu

## Abstract

Sewage sludge, a biosolid product of wastewater processing, is an often-overlooked source of rich organic waste. Hydrothermal processing (HTP), which uses heat and pressure to convert biomass into various solid, liquid, and gaseous products, has shown promise in converting sewage sludge into new materials with potential application in biofuels, asphalt binders, and bioplastics. In this study we focus on hydrochar, the carbonaceous HTP solid phase, and investigate its use as a bio-based filler in additive manufacturing technologies. We explore the impact of HTP and subsequent thermal activation on chemical and structural properties of sewage sludge and discuss the role of atypical metallic and metalloid dopants in organic material processing. In additive manufacturing composites, although the addition of hydrochar generally decreases mechanical performance, we show that toughness and strain can be recovered with hierarchical microstructures, much like biological materials that achieve outstanding properties by architecting relatively weak building blocks.

# I. Background

Sewage sludge, the biosolid product of wastewater processing, is an abundant source of rich organic compounds and nutrients including nitrogen and phosphorus. Each year the US generates 6 million dry metric tons of treated sewage sludge[1], and globally this figure is estimated to be around 144 million dry metric tons[2]. Treated sewage sludge can be used in land applications such as soil amendment or fertilizer. However, presence of microplastics and other pollutants limit land distribution, and over half of the sewage sludge produced in the US is incinerated, sent to landfills, or otherwise disposed of, escalating issues associated with greenhouse gas emissions and solid waste accumulation[1]. There is a clear need for better solutions to the prevalence of sewage sludge.

In typical wastewater processing, primary sludge is extracted with chemical precipitation and sedimentation, and suspended organic matter is extracted with biological treatment as secondary sludge, after which the sludges are combined for further treatment. At this stage, anaerobic or aerobic digestion is used to stabilize the solids and neutralize harmful pathogens. After the sludge is dewatered to further concentrate the solid matter, it is ready for reuse or disposal[3].

Treated sewage sludge has been investigated for use as toxic compound adsorbents[4–8], as cement fillers for structural materials[9–11], and in other applications[12,13], both as-is and after further processing. In particular, a promising method for valorizing sewage sludge is hydrothermal liquefaction (HTP), a process that uses high heat and high pressure to convert wet biomass into crude-like bio-oils and solid residues known as hydrochar, as well as some gaseous products (**Fig. 1a**)[14–16]. The yield and chemical formulation of HTP products depends highly on biomass source and reaction conditions. HTP is often preferable to pyrolysis for processing biomass because it does not require pretreatment and drying.

Hydrochar produced through HTP can be further processed into activated carbon through physical activation with heat, or through chemical activation with additional catalysts. The activation process generally decreases heteroatom content and increases carbon crystallinity within the solid, which can imbue additional properties such as enhanced strength or conductivity. Hydrochars and activated carbons (ACs) from lignocellulosic fibers, crustacean shells, and other biomass sources have been investigated as additives in 3D printing, often by incorporation into polylactic acid (PLA)[17–19], as well as for other composites to reduce raw material consumption and to lend additional functionality[14,15,20–22]. HTP for sewage sludge management has been found to eliminate toxic chemicals and to have an 11-fold higher energy recovery than landfilling sewage sludge[23,24]. Some studies have further investigated effects of different parameters on sewage sludge HTP yields[25–27], however most focus on the production of fuels including solids for char combustion[28], liquid bio-fuels[28,29], or methane-rich biogas[30].

Here we investigate a different route for the valorization of sewage sludge hydrochar to enhance the sustainability of 3D printing materials. Especially considered as a byproduct of biofuel production, utilizing sewage sludge in printed composites reduces consumption of raw synthetic materials and enables more sustainable waste management. We explore the impact of temperature and pressure on bio-oil and hydrochar yield and composition, as well as changes effected by physical activation. We further investigate the printability and mechanical characteristics of hydrochar-resin composites. Finally, we explore how nature-inspired strategies such as hierarchical material architecting can improve functionality of the composites, even when increasing sustainability comes with a trade-off of mechanical properties.

## II. Results

### 2.1 Hydrothermal liquefaction of sewage sludge

Sewage sludge was subjected to hydrothermal processing under five different conditions that are typical for biomass HTP: 250°C, 300°C, and 320°C at 0 bar, as well as 320°C at 10 bar and 20 bar[31,32]. Results are summarized in **Fig. 1b** and **Table S1**. Products of HTP are governed by a balance between various parameters including feedstock, loading, temperature, pressure, and time. Here, as a whole, yield of biocrude oil increased as temperature increased while yield of hydrochar decreased. This is consistent with previous studies that indicated the optimal HTP conditions for biocrude oil production from sewage sludge to be approximately 320°C or 330°C[29]. In similar vein, high temperature (>260C) HTP is often termed hydrothermal liquefaction (HTL) for its tendency toward high biocrude yield[23]. Conversely, yield of biocrude oil decreased and yield of biochar increased as pressure increased. High pressure may have been favorable for repolymerization reactions that enhance the formation of hydrochar[33]. The aqueous and gaseous phases were not captured. High temperature and low pressure (0 bar, 320°C) resulted in the highest biocrude oil yield, which is desirable for common biofuel and asphalt applications, while low temperature and low pressure (0 bar, 250°C) resulted in the highest hydrochar yield.

### 2.2 Physical & chemical characterization

The subset of hydrochars produced at 0 bar and 250°C, 300°C, and 320°C, which we will reference as H250, H300, and H320, were selected for further analysis since these lower pressure operating conditions are safer and more energy efficient and offered comparable or higher biocrude and hydrochar yields than their high-pressure counterparts. For further valorization, these hydrochars were subjected to thermal activation to produce activated carbon (AC), a process that typically produces porous carbonaceous materials with enhanced mechanical properties[34]. We will reference the respective activated carbons as AC250, AC300, and AC320 (**Table 1**).

Visually, thermal activation transformed the hydrochars from a dark brown color to a deep black color. SEM and XPS analysis were used to further consider the differences between activated and unactivated samples. SEM micrographs of H250 and AC250 are shown in **Fig. 2** as representative samples; micrographs of H300 and H320 and their respective ACs displayed similar trends and can be found in supplementary materials. Interestingly, powdered ACs appeared to have a higher quantity of large particles (>50 μm) than powdered hydrochars even though all samples underwent the same milling process. Upon closer inspection (**Fig. 2 c,d**), the hydrochars and ACs appeared to have similar microstructural topology and coarseness even though activation processes typically create more porous materials with increased carbon content [35]. At very high magnification, however, some nanostructural differences between hydrochars and ACs were revealed (**Fig. 2e,f**). In hydrochars, material surfaces were marred by networks of very fine nanoscale cracks on the order of 10s of nanometers. In ACs, surfaces were relatively smooth and carbon nanospheres could be found, demonstrating evidence of restructuring within the material.

XPS analysis of elemental composition revealed that carbon content universally decreased with thermal activation, the opposite effect of what is typically expected (**Fig. 3a**, **Table 2**). This may be attributable to the large presence of metallic and metalloid elements such as silicon and aluminum that are not typically found in biomass residues. Interestingly, one study also found that carbon is preferably transferred to the liquid phase in HTP processing of sewage sludge[36]. Here, with thermal activation, nitrogen content decreased as expected but all other dopants increased in relative concentration, likely because they could not be pyrolyzed while carbon and nitrogen-based compounds were. Furthermore, EDS scans showed that some dopants including aluminum and silicon were present as small grains rather than dispersed homogeneously throughout the material, and these increased in grain size after activation as if annealed, contributing to the larger particles observed in ACs. This is shown for H250 and AC250 in **Fig. 3b**, and EDS scans for the remaining materials can be found in supplementary materials.

Raman spectroscopy was used to assess the aromatic structure of the materials revealed structural reordering further: spectra of disordered graphitic carbons typically exhibit two bands around 1357 and 1580 cm$^{-1}$, often referred to as the D and G bands respectively, whose intensity and line width are characteristic of the carbon structure[37]. In particular, higher D band intensity is indicative of higher disorder including defects and vacancies, while higher G band intensity is representative of in-plane vibrations of aromatic sp$^2$ atoms[15]. Here, the D/G band intensity ratio was higher among the ACs than the hydrochars, indicating that the material had become more disordered after physical activation (**Fig. 3c**, **Table 3**). This suggests that despite some congregating of the inorganic dopants, the increased relative ratio of dopants still had a negative overall effect on graphitic ordering of the carbons. As before, this is contrary to what is typically expected for activation processes. Amongst the hydrochars, Raman spectra were internally consistent and displayed similar trends, as did the Raman spectra for the ACs. Nevertheless, both hydrochars and ACs displayed two relatively broad peaks characteristic of highly disordered carbons[37]. **Fig. 3** shows the elemental composition, EDS scans, and Raman spectra for H250 and AC250 as representative analysis; this data for the remaining samples can be found in supplementary materials.

The hydrochars were slightly more insulating than the ACs as evidenced by both slight shifting in their XPS spectra and by charge accumulation during SEM imaging. However, measurable conductivity was not detected in any of the samples with resistivity measurements on pressed powders, powders dispersed in deionized water, or in hydrochar and AC-doped composites. Some bio-based activated carbons do demonstrate conductivity[14,15], however conductive ACs typically have much higher carbon content (90+ wt%) than the sewage sludge-derived ACs here (50-60 wt%), which enables higher carbon crystallinity and electron delocalization[38].

## 2.3 Incorporation of sewage sludge into 3D printing materials

So far, we have investigated the products of hydrothermal processing of sewage sludge and the effects of an additional thermal activation step. As a potential route for valorization, we next investigated the use of the resulting sewage sludge hydrochars and ACs as sustainable fillers for 3D printing. Briefly, for each sample, powdered hydrochar or AC was mechanically mixed into liquid-crystal display (LCD) 3D printing resin and sonicated for dispersion, and this mixture was used directly in 3D prints (**Fig. 4a**). These materials were tested with microindentation and in tension and compression to determine bulk mechanical properties of the resulting composites.

Microindentation revealed that both indentation hardness and indentation modulus decreased with incorporation of hydrochar, and even more so with incorporation of ACs (**Fig. 4b,c**). As shown in **Fig. 4d-f**, tensile modulus, tensile toughness, and compressive modulus all demonstrated similar trends. This is consistent with some findings that incorporation of various hydrochars and ACs into 3D printing materials like PLA resulted in decreased tensile strength[39,40]. In contrast, some studies have found that using such fillers increased tensile modulus in their composites[39–43]. This has been proposed to be a result of the higher rigidity of hydrochar and AC fillers relative to 3D plastics[44], and the high surface area of the biochar which enhances adhesion between matrix and filler materials[45,46]. In this case, diminishing mechanical properties with the addition of hydrochar and AC suggests limited adhesion between the cured resin and the fillers, which inhibits stress transfer and allows deformability, or poor mechanical properties in the hydrochars and ACs themselves. In similar vein, elongation at break decreased with addition of the hydrochars and ACs, as is often seen in plastics with nonintercalated fillers, which can aggregate and cause embrittlement[47]. Nevertheless, up to 5 wt% hydrochar was still easily incorporable and printable (**Fig. 4g**), demonstrating that significant portions of synthetic 3D-printing resins can be replaced with sewage sludge as a more sustainable filler, especially in applications such as visual prototyping where mechanical strength is not of pinnacle importance.

An interesting observation is that the hydrochar-filled composites displayed stronger mechanical properties than the AC-filled composites. One potential explanation for this is the nanoscale cracks previously observed throughout the hydrochars, which may enhance adhesion with resin by increasing surface area and promoting resin impregnation. Another potential explanation is the higher degree of order observed in hydrochars, which may contribute to stronger mechanical properties within the hydrochar powders themselves as compared to the ACs.

Because the mechanical behavior of composites with the various hydrochars were quite similar, H250 was selected as a representative material for the remaining experiments based on the quantity of material available. The hydrochar imbued a rich dark color to printed specimens, as can be seen in **Fig. 5a**.

## 2.4 Use of gyroid geometries to "recover" mechanical properties

Many materials in nature are composed of relatively weak building blocks, but achieve remarkable properties through careful architecting of their microstructures[48–51]. With this inspiration, we investigated the use of geometry to enhance the mechanical properties of the relatively weaker hydrochar composites. A bio-inspired gyroid geometry was employed, which is trigonometrically approximated by **Eq. (1).**

$$sin(x)cos(y) + sin(y)cos(z) + sin(z)cos(x) = 0 \qquad (1)$$

Gyroids are triply periodic minimal surfaces that have been observed in butterfly wing scales, bird feathers, and mitochondrial membranes[52,53]. As a lattice structure, gyroids have been shown to yield high tensile strength materials with low density as well as near isotropy, and have been investigated in the design of high-performance carbon-based materials[54–56]. Therefore, we expect that in lightweight composites, we can use the gyroid structure to compensate for the decreased mechanical properties resulting from sewage sludge incorporation. Interestingly, due to the intricate pattern of connected surfaces and passages, gyroid geometries were near-impossible to manufacture until the advent of 3D printing.

We contrast hydrochar composite printed in gyroid geometries with resin alone printed in a traditional rectilinear lattice geometry. Gyroid architectures were also printed with pure resin for comparison. Printed samples, and their respective unit cells, are shown in **Fig. 5a-c**. With matched unit sizes, we found that across several relative densities, the rectilinear lattice with neat resin demonstrated the highest modulus and tensile strength (**Fig. 5d,e**). However, both the gyroid prints with neat resin and hydrochar composite surpassed the rectilinear lattice in shear strain and toughness up to relative density of approximately 0.3. The neat resin gyroids generally performed better than the hydrochar gyroids as expected. Nevertheless, this demonstrates that geometry can be used to compensate for some material weaknesses when using more sustainable materials, such as in packaging or similar applications where lightweighting and energy absorption may be more important than stiffness.

## III. Discussion and conclusions

This study investigated the valorization of sewage sludge with hydrothermal processing and additive manufacturing. A key observation was that the significant presence of metallic and metalloid dopants in sewage sludge, which are not typically found in biomass residues, were retained throughout the hydrothermal process and had a profound multiscale effect on the results of thermal activation to yield materials that are atypical of these processes. Many HTP studies explore the impact of varying HTP conditions on the properties of resulting materials, but here it was observed that the different HTP products varied minimally. Hydrothermal processing of sewage sludge created highly textured hydrochars with coarse microstructure and nanoscale cracks throughout (**Fig. 2**). Interestingly, dopants of different elements were not spread homogenously throughout the material but tended to congregate in individual particles. Thermal activation typically yields increased porosity, increased carbon content, and enhanced mechanical properties[15,57]. However, in this case, the dopants likely interfered with typical carbonization reactions and were not pyrolyzable, so thermal activation decreased the relative carbon content in the resulting activated carbons (**Fig. 3**). Furthermore, the activated carbons demonstrated decreased carbon ordering and aromaticity, smoother nanoscale topology without the presence of cracks, and larger particles with dopant impurities.

As a filler in resin-based additive manufacturing, we found that both hydrochars and activated carbons decreased the stiffness and hardness of composites, which may be attributable to limited adhesion between the printing resin and sewage sludge derivatives. Interestingly, the hydrochar-doped composites performed better than the activated carbon-doped composites, which may result from smaller particle sizes and better resin impregnation due to nanoscale cracks throughout the materials.

These structural and mechanical results present new phenomena contrary to typical understanding of biomass HTP and thermal activation, demonstrating that sewage sludge may not be able to be treated like typical biomass residues. This presents potential drawbacks, as much more analysis of sewage-sludge based materials is necessary, as well as opportunities, as new functionalities can be discovered.

Here, we further demonstrated that despite decreased stiffness, hydrochar-doped composites can outperform pure synthetic resin in toughness and failure strain by architecting it in a nature-inspired gyroid geometry, as compared to traditional rectilinear lattices. This phenomenon demonstrates a merger of geometry and material that is common to hierarchical biological materials, and shows that sewage sludge can still be successfully used as a sustainable filler to decrease consumption of raw synthetic material. Some potential applications include visual prototyping and packaging, *i.e.* applications where toughness may be more important than modulus or strength alone. Furthermore, we conclude that sewage sludge hydrochar may not be the best option as a 3D-printing additive, but it is still functional and a good outlet for excess residues that may result from sewage sludge HTP for other applications.

In our perspective, it would perhaps be more interesting to further investigate the effects of various dopants on HTP and carbon activation, such as how results would vary with different methods of activation. Mixing sewage sludge with other biomass waste streams could also potentially dilute the dopants enough to enhance carbon ordering or to harness other properties such as conductivity. Understanding how dopants affect functional bio-based materials may have implications on the valorization of various biomass residues, especially as the field of sustainability shifts towards using waste streams that may not have tightly controlled constituents. In fact, it is possible that these dopants may have properties that can yield new functionalities to biobased materials, such as in one study that found that arsenic removal was positively correlated with inorganic content in carbonized waste materials for water treatment[34]. We present a first foray here and anticipate further research on the valorization of disordered biomass waste streams.

## IV. Methods

### 4.1 Hydrothermal processing

Sewage sludge was procured from Bay State Fertilizer (Quincy, MA), which dewaters and pelletizes digested sewage sludge from the Deer Island wastewater treatment plant (Winthrop, MA). The digested sludge is composed of 60-65% primary sludge by weight, with the remainder being secondary sludge.

Batch HTP experiments were carried out with a 1.8 L Parr reactor (Parr Instrument Company, Mode 4578). To the 1.8 L chamber, 200 g (dry weight) of sewage sludge was loaded, and D.I. water was added to reach a solid content of 20% (w/w). The reactor was tigHTPy sealed and pressurized with pure nitrogen to reach the target initial pressure (0, 10, or 20 bar). The reactor was heated up to the target temperature (250, 300, or 320 ºC) at a rate of 3.33 ºC/min, then held at the target temperature for 30 minutes before cooling down. When the temperature of the reactor dropped below 50 ºC, the residual pressure and temperature were recorded before releasing the residual pressure and opening the chamber. The resulting slurry was transferred to a 2 L container, and a small amount of water was used to wash the chamber to recover all slurry. To separate the products, 400 mL dichloromethane (DCM) was added to the slurry and mixed well. The mixture was then filtered through a Büchner funnel with one Whatman GF/B glass filter (diameter 110 mm) with vacuum. The solid cake in the Büchner funnel was rinsed again with 400 mL DCM, then dried at 105 °C overnight to yield the hydrochar. The liquid DCM layer was separated from the aqueous layer, then dried at room temperature in a fume hood to remove the DCM until mass of the mixture plateaued. Finally, the resulting bio-oil was dried at 60 °C for 1 hour to remove residual DCM.

Hydrochar and bio-oil yield were calculated as the ratio of the weight of the recovered product mass ($m_i$, where $i$ = hydrochar, or bio-oil) and the dry mass of sewage sludge ($m_{sludge}$) initially loaded to the reactor, according to **Eq. (2)**:

$$Yield\ (wt\%) = \frac{m_i}{m_{sludge}} \cdot 100 \tag{2}$$

## 4.2 Thermal activation & ball milling

Thermal activation of hydrochar was conducted in a Carbolite Gero tube furnace under continuous flow of nitrogen. For each sample, 4g of hydrochar was placed in a ceramic crucible. Temperature was ramped from room temperature to 850°C at a rate of 5°C per minute, then held at 850°C for 2 hours, after which the furnace was allowed to cool to room temperature before opening. Activated and non-activated samples were ground with a ball-mill (SPEX SamplePrep, 8000M Mixer/Mill) for 5 minutes to obtain a fine powder.

## 4.3 Physical & chemical characterization

X-ray photoelectron spectroscopy (XPS) was performed on a PHI VersaProbe II to evaluate elemental composition and surface chemistry of samples. Milled samples were pressed onto copper tape with a metal spatula, after which excess was tapped off, and mounted on a sample holder. X-rays were generated with a monochromatic aluminum Kα source. Pass energy of 187.85 eV and a 200 μm spot size were used, with a step size of 0.100 eV for high resolution scans and 0.800 eV for full sweeps. High resolution spectra were taken for C1s, O1s, and N1s. Data analysis was performed using CasaPXS (Neal Fairley).

Raman spectroscopy was performed with a confocal Raman microscope (Alpha 300Ra; WITec, Germany) using a Nd:YAG laser (λ=532 nm) and a 50x Zeiss objective. Spectra were acquired from ball milled HC and AC placed on glass microscope slides. The excitation wavelength was calibrated using a silicon wafer standard. For each sample, 10 individual Raman spectra were acquired at random locations, with an accumulation time of 20 seconds per data point and a laser power of 3-4 mW. After background removal, Lorentzian distribution functions were fitted to the spectra to evaluate the D-to-G band intensity ratio ($I_D/I_G$) based on the integrated areas. The average crystal planar domain size $L_a$ was then calculated using the relation formulated by Knight and White[37] and based on the work of Tuinstra and Koenig[58], using the coefficient 4.4 for the green laser as shown in **Eq. (3)**:

$$L_a = \frac{4.4}{I_D/I_G} \tag{3}$$

SEM was performed using a Vega3 XMU (Tescan, Czech Republic) SEM. This was paired with a Bruker Xflash 630 silicon drift detector for EDS elemental mapping. High-resuolution SEMS were acquired with a Zeiss Merlin High-resolution SEM. Hydrochar samples were gold-sputtered with a Denton Vacuum Desk V (Moorestown, NJ, USA) for 30 seconds before imaging.

## 4.4 Geometry design

Sample geometries and 3D printing STL files were prepared using nTopology software (nTopology Inc.)[59]. Compression sample geometries were 12.7mm (diameter) x 25.4mm (height) cylinders as designated in ASTM d695, and tensile sample geometries were taken from ASTM d638 Type IV. Bioinspired geometries were 12.7mm x 12.7mm x 25.4mm rectangular prisms, and were designed such that gyroid and rectilinear lattices had the same unit cell size, but varying wall thickness to achieve similar relative densities at 8 different points ranging from approximately 0.15 to 0.55. Samples for testing of bulk material properties were prepared at least in triplicate to assess reproducibility.

## 4.5 Resin incorporation & 3D printing

Milled hydrochar and AC samples were mechanically incorporated into Anycubic Plant-based UV photocurable resin. Hydrochar was incorporated at 2% weight/weight with constant stirring at 450 rpm for 30 minutes. The mixture was then sonicated for 5 minutes to ensure particle dispersion, and stirred for an additional 30 seconds to reduce bubbles.

Stereolithography 3D printing was executed with a mono LCD printer (Elegoo Mars 2 Pro), where resin is selectively cured layer-by-layer with an LCD screen. After printing, samples were removed from the build plate with a metal spatula, washed of excess resin for 2 minutes in an ethanol bath, and crosslinked with UV light exposure for 3 minutes (Elegoo Mercury Plus V2).

After printing, conductivity of pressed powders, powders dispersed in deionized water, and bulk printed materials was assessed with a multimeter.

## 4.6 Mechanical testing

Micro-indentation was performed using an Anton Paar Instruments system. All measurements were conducted with a Berkovich tip. 2.5 cm x 2.5 cm cylindrical samples were 3D printed, and sample surfaces were polished prior to testing using a sequence of polishing pads with reduced abrasiveness. Indentation hardness and indentation modulus were calculated for all data sets using the Oliver-Pharr method[60]. The specimens were loaded in force-control mode up to the maximum indentation depth limit of 24 μm at a loading rate of 3 N/min. At the peak load, the force was held constant for 10 seconds before unloading at a rate of 3 N/min. Each sample underwent 20 to 25 microindentation measurements.

Compression testing was performed on an Instron universal tensile testing system with a 100 kN load cell and 15 cm platens. Tensile testing was performed on an Instron universal tensile testing system with a 5 kN load cell and pneumatic grips.

**Acknowledgements**
NSF GRFP (grant #1745302)
Mathworks Engineering Fellowship Fund

**Author contributions**
Conceptualization: S.S., S.Z., A.M., M.B.

Data collection: S.S., B.S., S.Z., D. S.

Data curation and visualization: S.S., A.M.

Writing – original draft: S.S., B.S., S.Z.

Writing – review & editing: S.S., M.B.

**Competing interests**
The authors declare no competing interests.

**Data Availability**
All relevant data are provided or available from the author upon request.

**Supplementary Information**

**Figure S1**. SEM micrographs of sewage sludge-based hydrochars and activated carbons.
**Figure S2**. High-resolution SEM micrographs of sewage sludge-based hydrochars and activated carbons at 20KX magnification.
**Figure S3**. High-resolution SEM micrographs of sewage sludge-based hydrochars and activated carbons at 40KX magnification.
**Figure S4**. Elemental analysis of sewage sludge-based hydrochars and activated carbons.
**Figure S5**. Physical and chemical analysis of sewage sludge-based hydrochars and activated carbons.

**Figures**

**a**

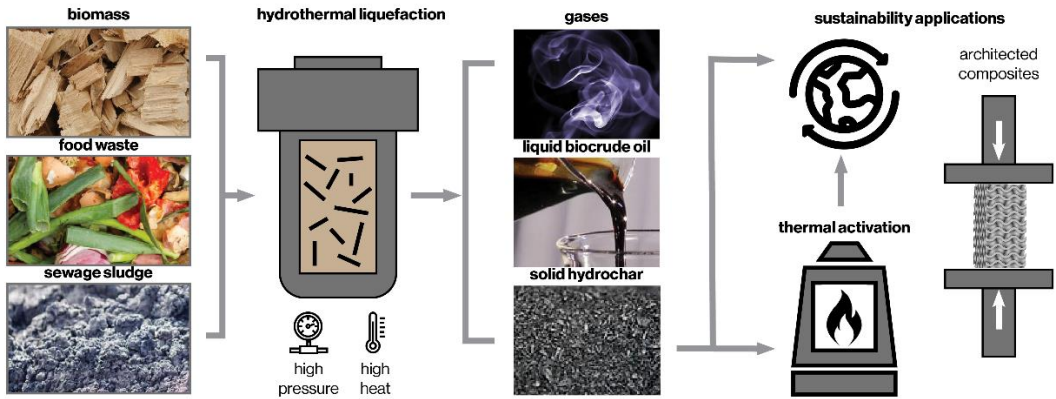

**b**

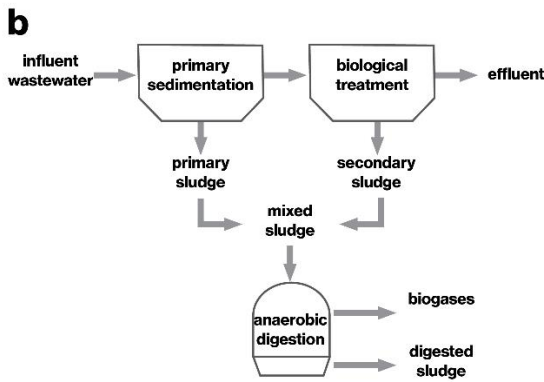

**c**

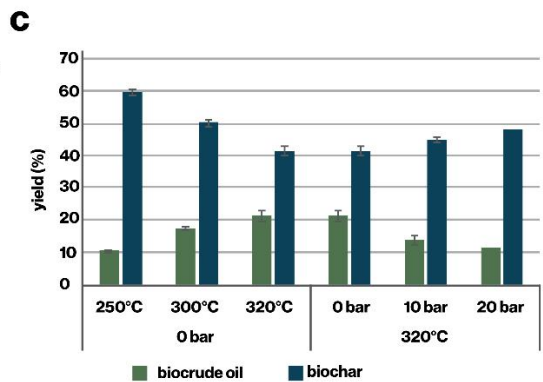

**Figure 1**. **Hydrothermal liquefaction of organic materials.** (a) Liquefaction of hydrated organic wastes under high heat and high pressure produces products in three phases: gases, liquid biocrude oils, and solid hydrochar components. Here, we investigate the valorization of solid hydrochar from sewage sludge, with and without further processing, in 3D-printing applications such as manufacturing of architected composites. (b) sewage sludge processing (c) Biocrude oil and hydrochar yields under several conditions of reactor pressure and temperature.

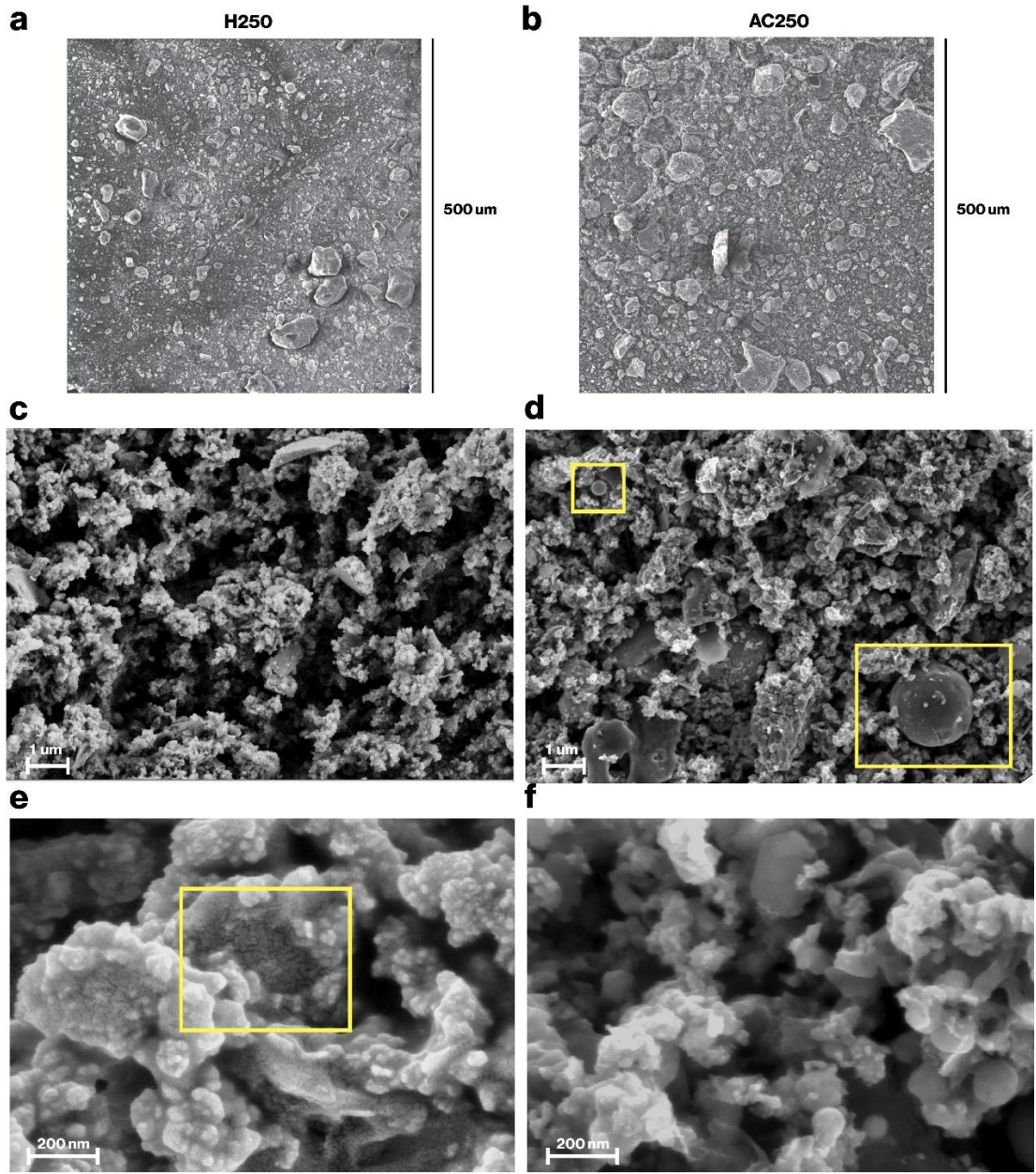

**Figure 2. Analysis of the multiscale structure of sewage sludge hydrochar (H250) and activated carbon (AC250), focusing on surface roughness and morphology** SEM micrographs of sewage sludge biochar directly after hydrothermal processing (a,c,e) and after further thermal activation (b,d,f). Hydrochar appears to have fewer large particles (a) than activated carbon (b). At 10 KX magnification, hydrochar (c) and activated carbon (d) have similar roughness and morphology, aside from the formation of carbon nanospheres in activated carbon. At 80 KX magnification, hydrochar exhibits a network of nanoscale cracks throughout its surfaces (e) while activated carbon appears to have sharper edges but smoother surfaces (f).

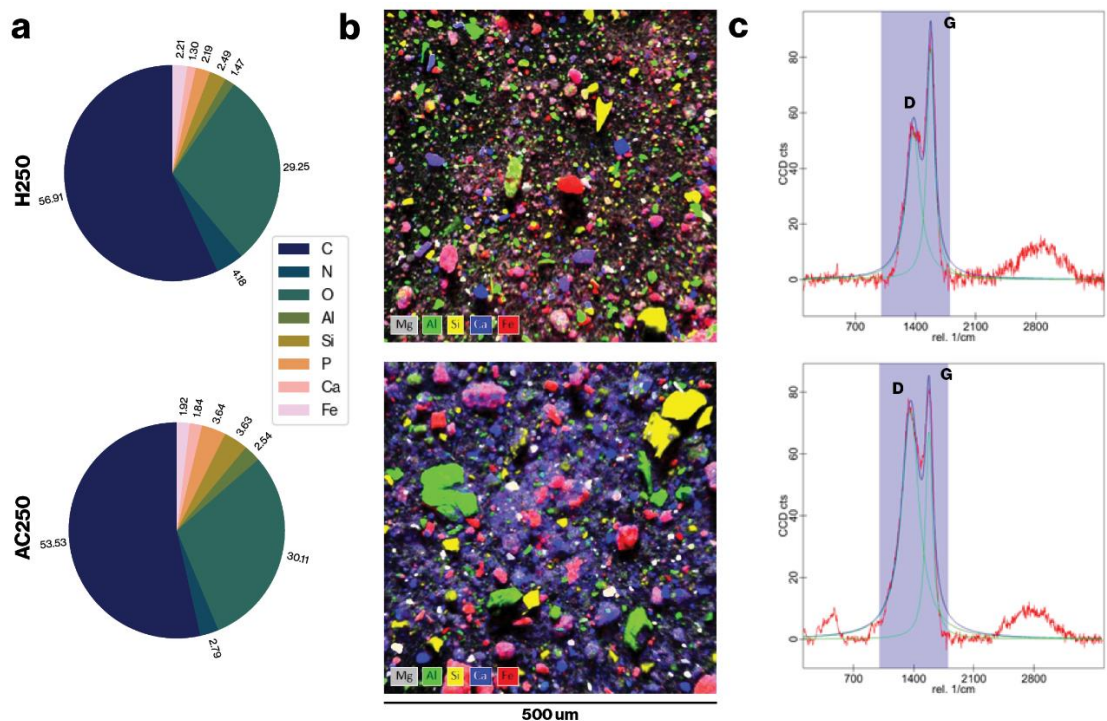

**Figure 3. Chemical and physical analysis of hydrochar and activated carbon.** Representative data from H250 and AC250; data from the remaining samples displayed similar trends and can be found in supplementary materials. (a) Elemental composition of hydrochar and activated carbon from XPS analysis. Relative carbon content decreased after thermal activation, while relative content of metallic and metalloid dopants increased. (b) Metallic and metalloid dopants were congregated in individual particles rather than dispersed throughout. Particle sizes in hydrochar (top) increased after thermal activation (bottom). Micrographs are of the same scale. (c) Raman spectra from hydrochar (top) and activated carbon (bottom) are characteristic of disordered carbons, with relatively broad D and G bands indicating higher disorder detected in the activated carbon.

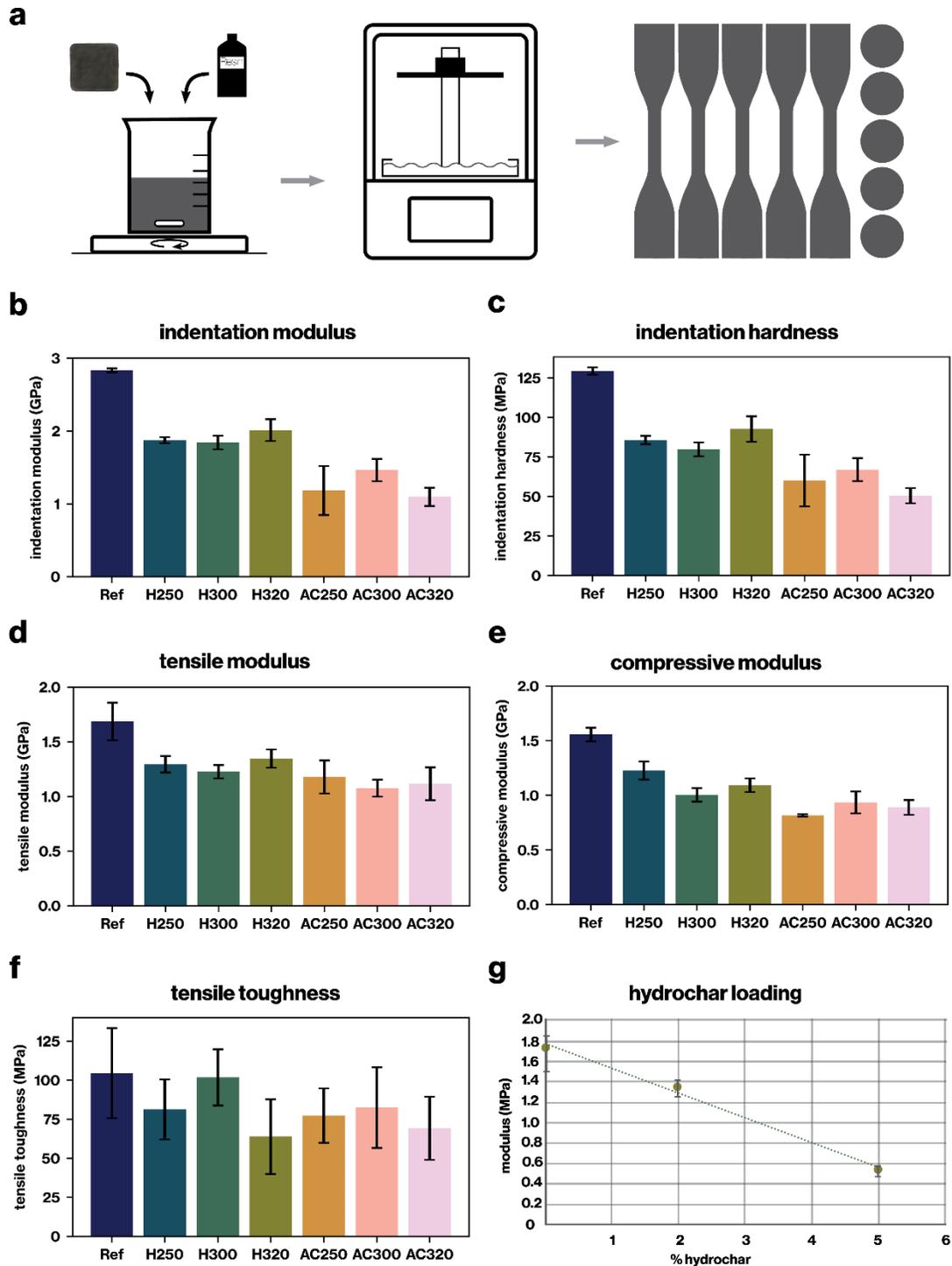

**Figure 4. 3D-printing with hydrochar-loaded resin, showing degradation of key mechanical properties with increasing hydrochar content.** (a) Hydrochar and commercial resin are combined with mechanical mixing and sonification for dispersal, after which the mixture can be used for 3D-printing with a DLP system. Mechanical testing samples in the shape of dogbones (tension, ASTM D638) and cylinders (compression, ASTM D695; microindentation) were fabricated with sample size n=5 for each condition. With microindentation, it is evident that incorporation of sewage sludge-based residues, especially activated carbons, decreases the

indentation modulus (b) and hardness (c) of printed composites. Similar trends are observed with tensile modulus (d), compressive modulus (e), and tensile toughness (f). (g) Composite modulus scales approximately linearly with respect to hydrochar (H320) loading. Nevertheless, mixtures of up to 5 wt% hydrochar can be successfully printed.

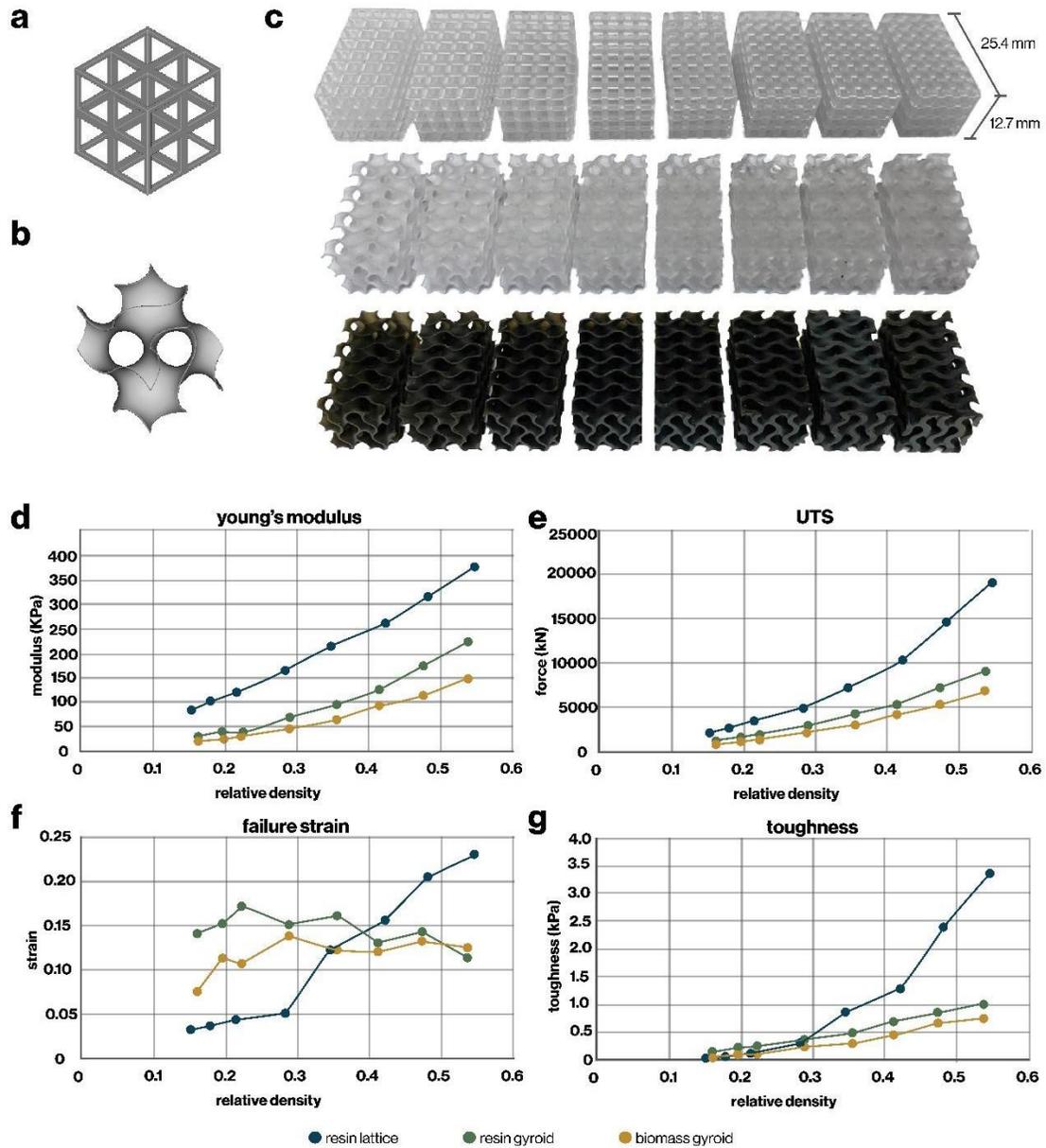

**Figure 5. Toughness and strain can be recovered at low density by architecting material structures.** Commercial resin was used to fabricate samples in a range of relative densities with (a) traditional lattice infills and (b) bioinspired gyroid infills. (c) Biochar-loaded resin was used to fabricate samples in a range of relative densities with bioinspired infills. These samples were tested in compression to find their (d) effective moduli, (e) fracture strength, (f) strain-at-failure, and, (g) toughness. Especially at lower densities, biomass-loaded resin can achieve toughness and elongation that exceeds conventional resin alone by using bio-inspired architecture.

**Table 1. Processing conditions for hydrochar and activated carbon samples.** Hydrochars were produced with hydrothermal processing at various temperatures and 0 bar, then converted into activated carbons with physical thermal activation.

| Sample | HTP Temperature (°C) | HTP Pressure (bar) | Thermal Activation |
|--------|----------------------|--------------------|--------------------|
| H250   | 250                  | 0                  | N                  |
| H300   | 300                  | 0                  | N                  |
| H320   | 320                  | 0                  | N                  |
| AC250  | 250                  | 0                  | Y                  |
| AC300  | 300                  | 0                  | Y                  |
| AC320  | 320                  | 0                  | Y                  |

**Table 2. Elemental composition of sewage sludge hydrochars and activated carbons.** In general, relative carbon content decreased with thermal activation while relative content of various heteroatoms increased, the opposite effect of what is usually observed with thermal activation of biomass-based hydrochars.

| Composition | H250  | H300  | H320  | AC250 | AC300 | AC320 |
|-------------|-------|-------|-------|-------|-------|-------|
| C           | 56.92 | 67.7  | 72.14 | 53.52 | 55.32 | 60.47 |
| N           | 4.18  | 4.3   | 3.61  | 2.79  | 1.45  | 2.37  |
| O           | 29.26 | 21.74 | 18.85 | 30.11 | 30.45 | 26.84 |
| Al          | 1.47  | 0.42  | 0.58  | 2.54  | 2.5   | 2.03  |
| Si          | 2.49  | 1.54  | 2.16  | 3.63  | 3.05  | 2.3   |
| P           | 2.19  | 1.91  | 1.15  | 3.64  | 3.27  | 2.67  |
| Ca          | 1.3   | 0.98  | 0.66  | 1.84  | 2.09  | 1.73  |
| Fe          | 2.21  | 1.42  | 0.85  | 1.92  | 1.88  | 1.59  |

**Table 3. The parameters of fitted Lorentzian functions to D and G bands: $x$ (mean value), $s$ (standard deviation), and $I_D/I_G$ (D to G band intensity ratio).** Based on the observed D to G band intensity ratios, the activated carbons were significantly more disordered in their molecular structures compared to the hydrochars, the opposite effect of what is usually observed with thermal activation of biomass-based hydrochars.

| Sample | $x_D$ [1/cm] | $x_G$ [1/cm] | $s_D$ [1/cm] | $s_G$ [1/cm] | $I_D/I_G$ [-] | $L_a$ [nm] |
|--------|--------------|--------------|--------------|--------------|---------------|------------|
| HC250  | 1374.9       | 1577.4       | 194.7        | 104.5        | 1.19          | 3.69       |
| HC300  | 1376.8       | 1572.6       | 181.6        | 105.6        | 0.93          | 4.72       |
| HC320  | 1373.9       | 1570.1       | 188.2        | 110.1        | 1.02          | 4.32       |
| AC250  | 1362.1       | 1577.6       | 245.7        | 98.6         | 2.75          | 1.60       |
| AC300  | 1359.5       | 1578.1       | 268.0        | 96.4         | 3.31          | 1.33       |
| AC320  | 1365.0       | 1576.0       | 258.6        | 93.8         | 3.37          | 1.31       |

Supplementary Materials

**Valorizing Sewage Sludge: Using Nature-Inspired Architecture to Overcome Intrinsic Weaknesses of Waste-Based Materials**

Sabrina C. Shen, Branden Spitzer, Damian Stefaniuk, Shengfei Zhou, Admir Masic, Markus J. Buehler[*]

[*]mbuehler@mit.edu

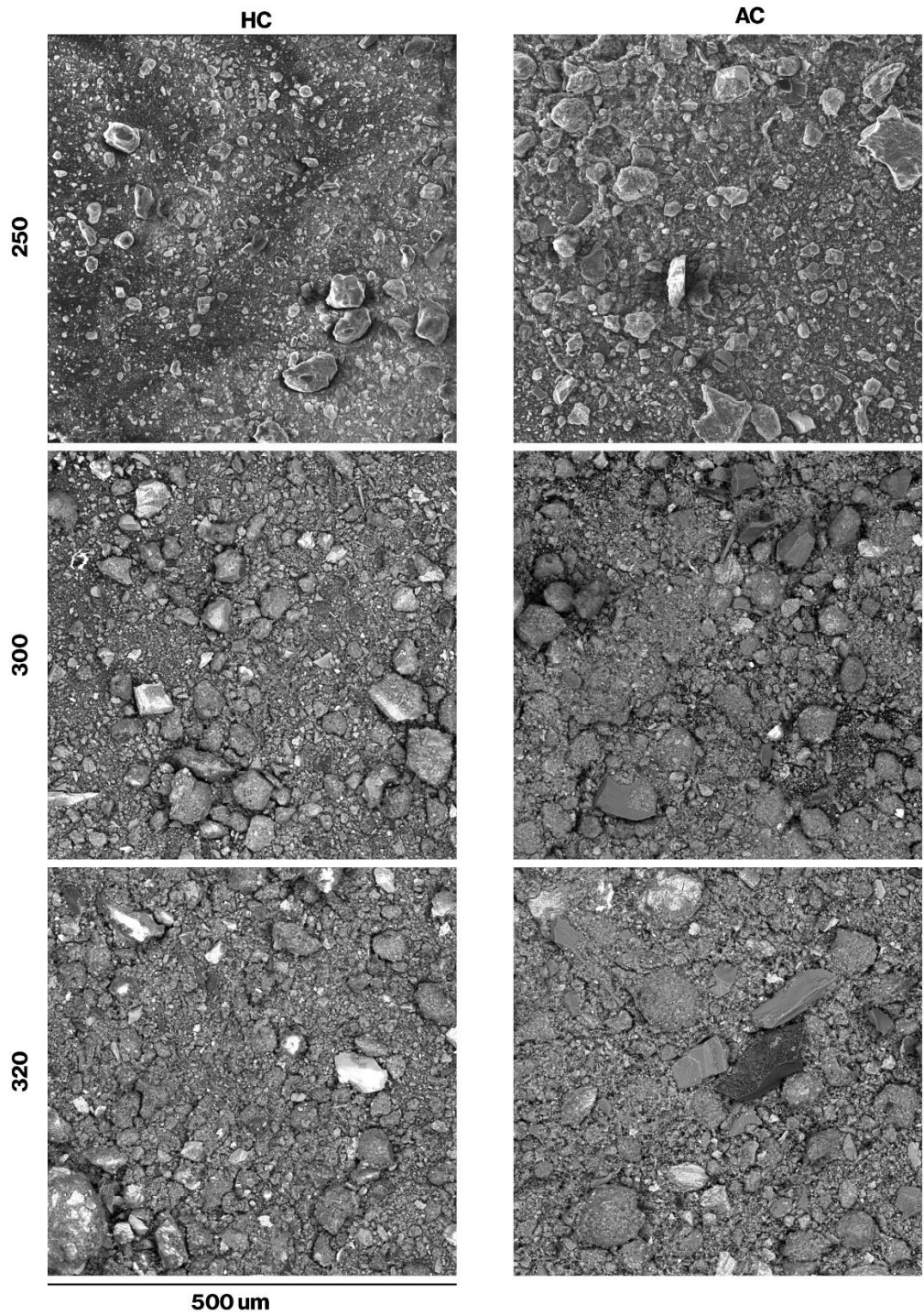

**Figure S1. SEM micrographs of sewage sludge-based hydrochars and activated carbons.**

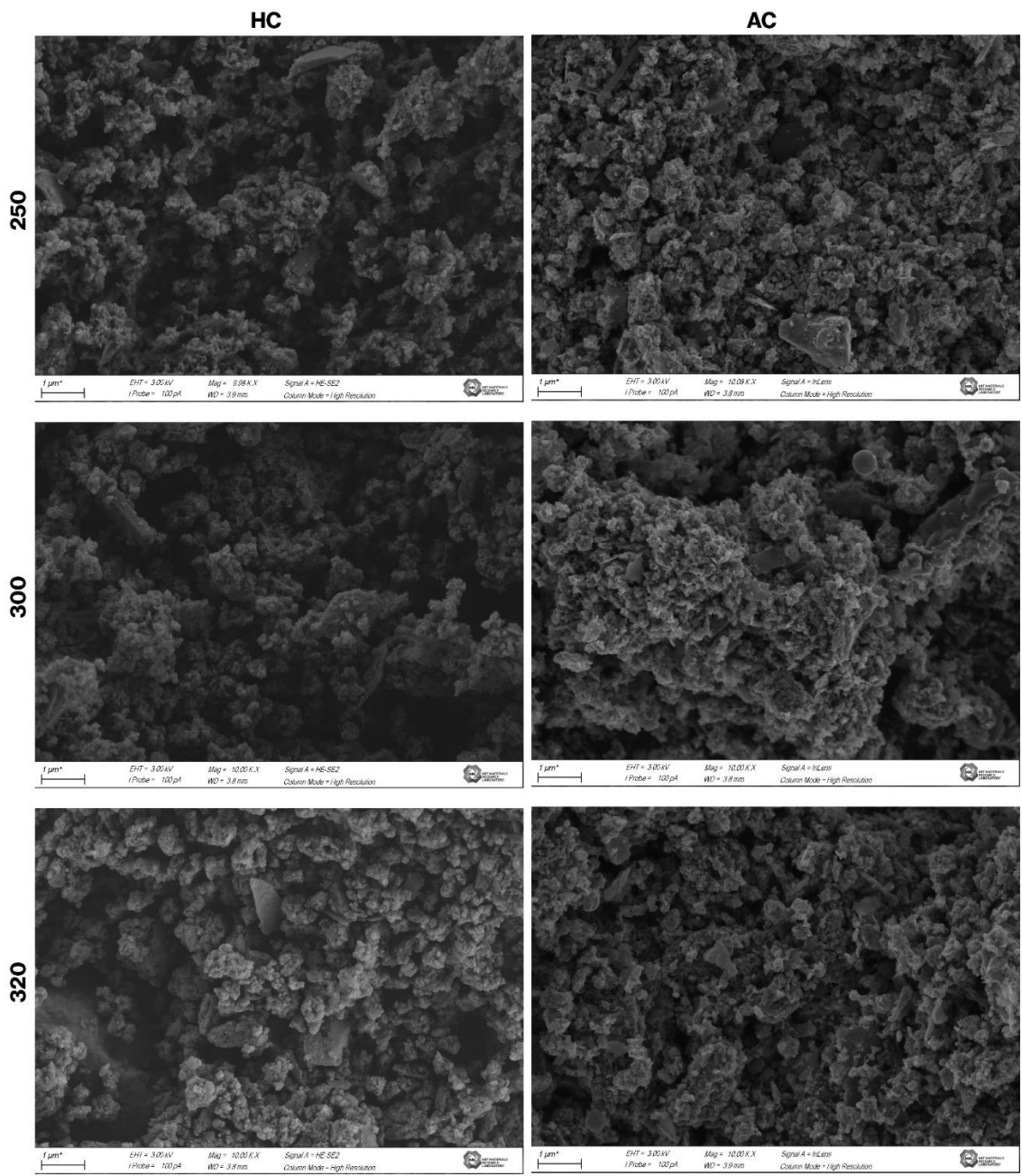

**Figure S2.** High-resolution SEM micrographs of sewage sludge-based hydrochars and activated carbons at 20KX magnification.

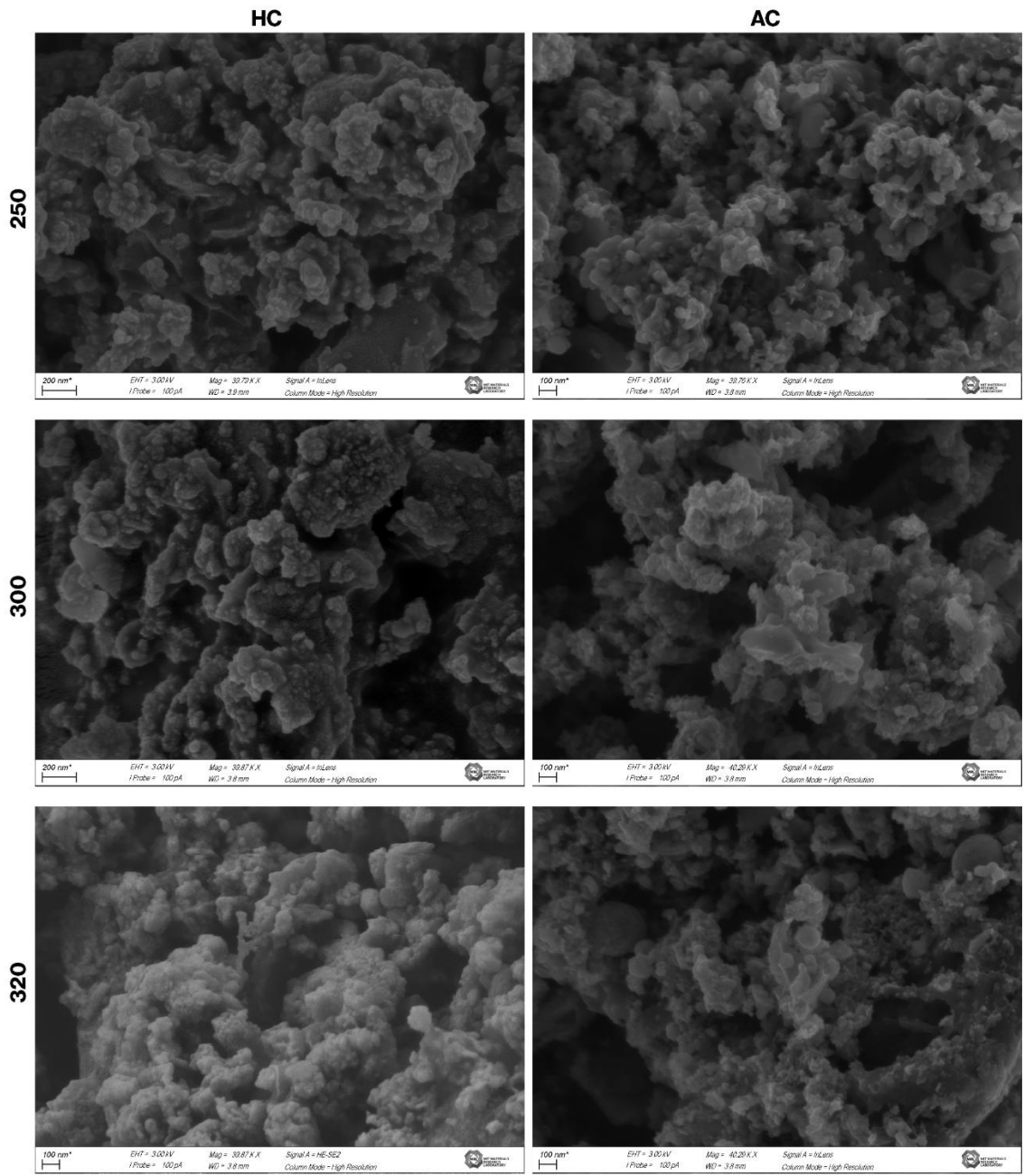

**Figure S3. High-resolution SEM micrographs of sewage sludge-based hydrochars and activated carbons at 40KX magnification.**

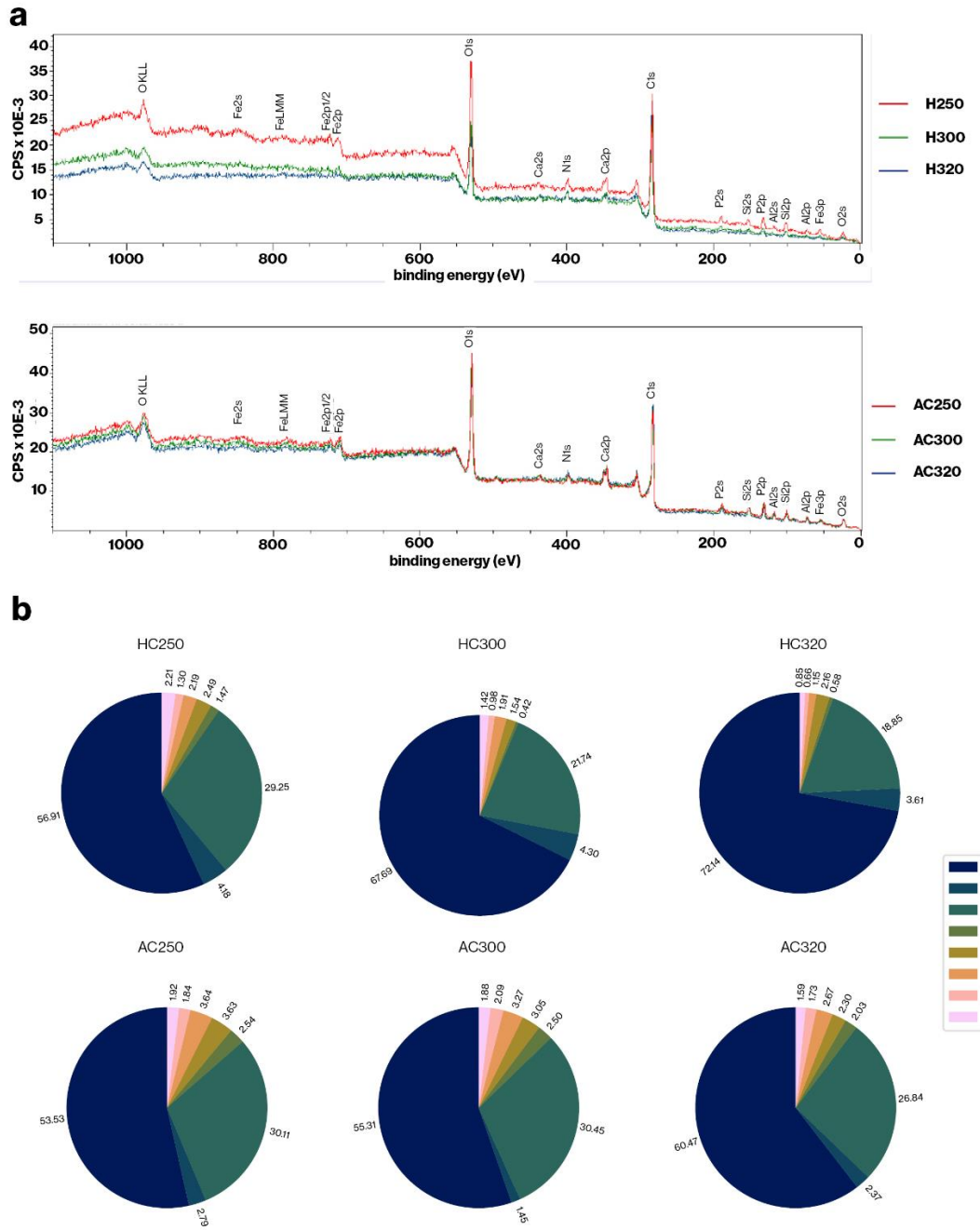

**Figure S4. Elemental analysis of sewage sludge-based hydrochars and activated carbons.** (a) XPS spectra of hydrochars (top) and activated carbons (bottom). (b) Elemental composition of hydrochars (top) and activated carbons (bottom).

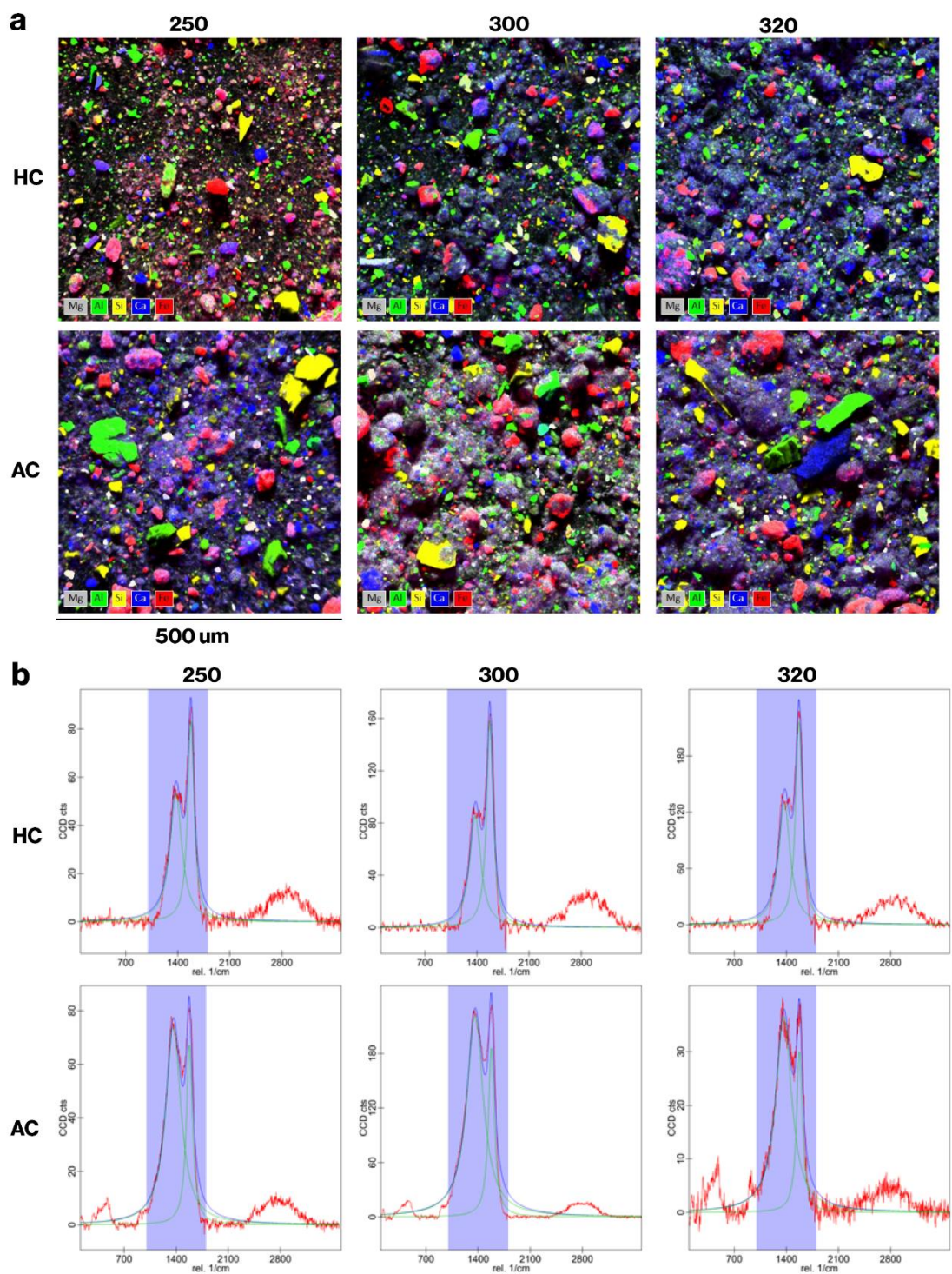

**Figure S5. Physical and chemical analysis of sewage sludge-based hydrochars and activated carbons.** (a) EDS scans of hydrochars (top) and activated carbons (bottom). (b) Raman spectra of hydrochars (top) and activated carbons (bottom).